\documentclass[preprint,12pt]{elsarticle}

\usepackage{graphicx}
\usepackage{caption}
\usepackage{subcaption}
\usepackage{amssymb}

\usepackage{lineno}

\begin{document}

\begin{frontmatter}


\title{Multi-Fidelity Digital Twins: a Means for Better Cyber-Physical Systems Testing?}



\author{Aitor Arrieta}

\address{Mondragon University, Spain}

\begin{abstract}

Context: Cyber-Physical Systems (CPSs) combine software and physical components. These systems are widely applied in society within many domains, including the automotive, aerospace, railway, etc. Testing these systems is extremely challenging, therefore, it has attracted significant attention from the research community. A driving CPS testing technique in industry is simulation-based testing. However, this poses significant challenges. 

Objectives: To propose a novel approach that deals with the CPS testing problems.

Method: In this new-idea paper we present the concept of multi-fidelity digital twins to deal with CPSs testing problems. 

Results: An initial hypothesis and open research questions are presented.

Conclusion: The paper presents the research foundation on the concept of multi-fidelity digital twins and provides a set of unexplored research areas.

\end{abstract}


\end{frontmatter}


\section{Introduction}

Cyber-Physical Systems (CPSs) combine digital cyber technologies (e.g., microprocessors and networks) with parallel physical processes \cite{derler2011modeling}. The dominant technology for testing these systems is simulation through digital twins. For over 20 years, the industries developing CPSs have relied on the well-known “X-in-the-Loop” simulation-based testing levels for testing embedded software. While these testing levels have are established in thousands of industrial companies for testing their systems, they pose several disadvantages. One such disadvantage is that executing tests is usually expensive, due to a variety of reasons \cite{arrieta2019pareto}: First, the physical layer of these systems is usually modelled by complex mathematical models. For instance, in the industrial example reported in \cite{gladisch2019experience}, when simulating autonomous vehicles, a simulation run is taken down to 10\% of real time factor, meaning that a simulation of 1 minute takes 10 minutes to execute. Secondly, tests need to be executed at different test levels (i.e., Model, Software and Hardware-in-the-Loop test level). In addition, a recent study also recommended executing the same test scenarios at the same test levels but combining different simulators as the test results might differ \cite{borg2020digital}, which would, again, increase the overall test execution time. Thirdly, as different engineering domains exists when developing CPSs, simulation is often turned into a co-simulation, which increments the execution time of test cases due to the need of intercommunicating between simulation tools. Lastly, test cases usually differ from those scenarios that are found in reality. To solve all these testing issues, recent approaches have proposed cost-effective methods for several testing stages, such as test generation \cite{menghi2020approximation,abdessalem2018testing} or test selection \cite{arrieta2019pareto}.

Because of the increase on both the complexity and the autonomy of CPSs (which increasingly incorporate Artificial Intelligence (AI) algorithms to control complex and often safety-critical tasks), limiting CPSs verification and validation tasks to these levels might be reckless. Furthermore, the software of CPSs has a long lifecycle (over 20 years in industries like railway and elevation \cite{ayerdi2020towards}. Subsequently, embedded software evolves while the system is in operation to deal with (1) legislation changes, (2) bugs found in production, (3) new functionalities demanded by the end user, (4) hardware obsolescence, etc. Already established technology, such as fast connection technology (e.g., 5G) and novel hardware platforms (e.g., mixed-criticality based embedded systems), allow for pushing the X-in-the-Loop simulation-based testing methods to the next level. In this paper we propose the concept of \textbf{multi-fidelity digital twins} and novel approaches that would help to significantly increase the cost-effectiveness of testing CPSs on the basis of the following hypothesis:

\noindent\fbox{%
    \parbox{\textwidth}{%
        \textbf{Hypothesis:} Multi-fidelity digital twins can significantly enhance the CPSs testing practices.
    }%
}

\section{Multi-Fidelity Digital Twins}

A digital twin is “a comprehensive digital representation of an individual product that will play an integral role in a fully digitalized product life cycle” \cite{haag2018digital}. When engineers develop CPSs digital twins, they make assumptions while pursuing the goal of abstracting the digital twin model at a level where testing is permitted. For instance, when developing a digital twin of an autonomous vehicle for testing its safety functionalities, it is possible to assume that its engine will always perform similar to reality by developing a simple mathematical model of the output torque produced by the engine given specific inputs. This way, engineers omit parts of the CPS that are out of the \textbf{scope of testing interest} for testing the safety functionalities of the autonomous car (e.g., temperature of the engine, revolutions, etc.). This will facilitate a significant reduction of testing time because a complex model of car engine could significantly increase the simulation time. Instead, engineers focus on developing the functionalities that autonomous vehicles have (e.g., line keeping, etc.).

\begin{figure*}[t]
	\centering
	\includegraphics[trim = 0 230 80 0,clip, width=1.15\linewidth]{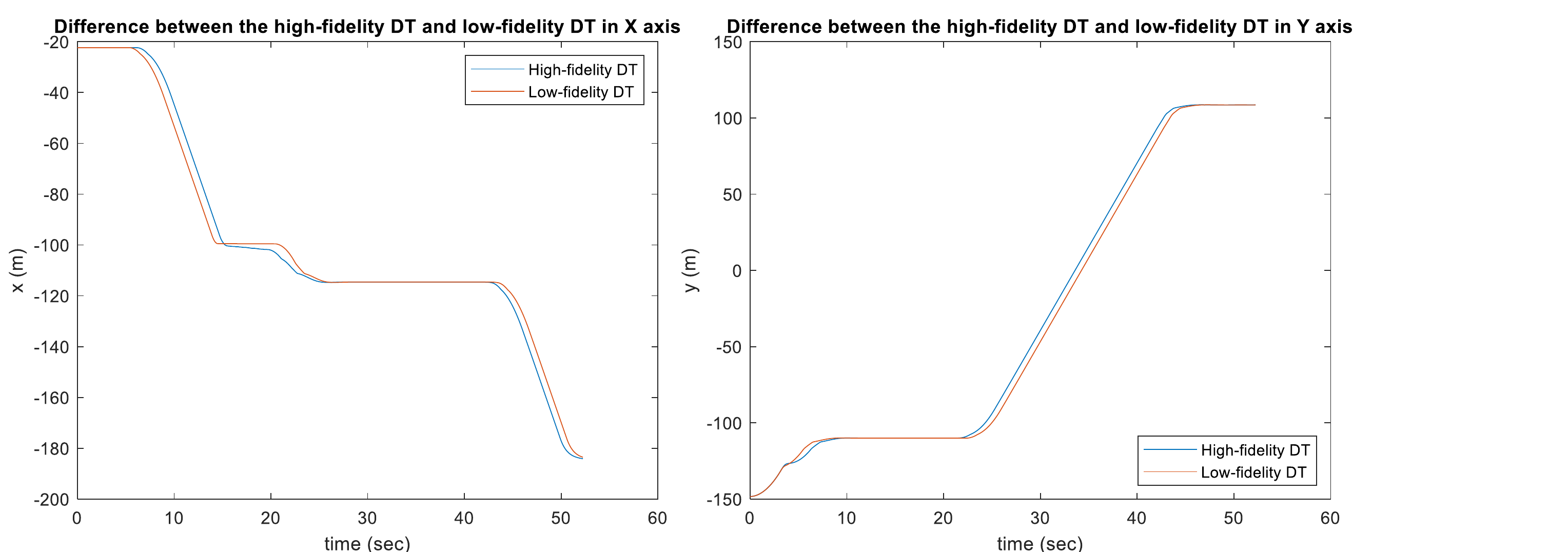} 
	\caption{Comparison of the simulation results in X and Y for the low-fidelity and high-fidelity digital twins}
	\label{fig:Approach}
	\hfill	
\end{figure*}

In a similar line, let us consider as an example of a preliminary experiment a digital twin of an autonomous car we modelled in MATLAB/Simulink. In such case, the automated driving toolbox developed by Mathworks provides some virtual sensors (e.g., cameras, radars, etc.) with the goal of detecting front vehicles. Suppose a new version of the lane keeping algorithm is being developed. Thus, the scope of testing interest is the subsystems implementing these functionalities, which includes front cameras, artificial vision algorithms, the lane keeping algorithm, etc. In this example, to illustrate the concept of fidelity, we have replaced the subsystem for object detection with a more simple algorithm making certain assumptions. For instance, instead of including the radar sensor for detecting objects, we process the position of different vehicles in the environment, and if any of these is close to collision, our simple algorithm triggers the flag for detecting the car. The test execution difference for a test scenario of 52 (simulated) seconds was 216 (real) seconds (i.e., 23.23 seconds for the simplified version and 239.15 seconds for the model that considers the virtual radar). This means that our assumptions for simplifying the collision detection algorithm reduced the test execution time by 90\%. For this scenario, which included the detection of a single vehicle, Figure \ref{fig:Approach} shows the difference between using the simplified version (low-fidelity digital twin) of the subsystem and the digital twin that includes the virtual sensor of the radar (high-fidelity digital twin). As can be seen, the differences are minimal.

With this assumption, we have significantly reduced the test execution time, increasing the possibilities of executing more tests for testing the new lane keeping control software version. However, we have also reduced the fidelity of the digital twin by substituting the radar-based object detection subsystem with a simpler algorithm that makes certain assumptions (e.g., the radar system will always provide the same information as the one processed by our (faster) algorithm based on the position of the vehicles). By having a repository of subsystems with assumptions like these ones that simplify the original digital twin, it could be possible to generate multiple digital twins of multiple fidelity levels, which can be used for testing purposes that are within several scopes of testing interests. We term these types of digital twins as “multi-fidelity digital twins”, and this fact leads us to the following open question, which it is important to answer before proposing this paradigm for CPSs testing:

\noindent\fbox{%
    \parbox{\textwidth}{%
        \textbf{Question 1:} How can the fidelity of a digital twin be quantitatively measured?
    }%
}

\

The fidelity of a digital twin has a direct relation with different properties. These can be based on the accuracy with respect to the physical twin (i.e., the real system), the digital twin trustworthiness, repeatability of the results, scope of the testing interest, etc. Depending on such characteristics, different ways of measuring the fidelity of a digital twin can be proposed, leading to several fidelity metrics. For many of them, data from the real CPS (i.e., the physical twin) will be necessary, which requires the system to be in operation. However, once this is solved, and having reliable fidelity metrics, it will be possible to automate several tasks in relation to the generation and maintenance of multi-fidelity digital twins, which leads us to the following open question:

\noindent\fbox{%
    \parbox{\textwidth}{%
        \textbf{Question 2:} How can multi-fidelity digital twins be (semi)-automatically generated and maintained?
    }%
}

The use of data from a CPS in operation can help measure the fidelity of a digital twin in many aspects (e.g., how accurate a digital twin is with respect to its physical twin). Furthermore, this data can also help on proposing changes in the digital twin to increase its fidelity by tuning certain parameters. To this end, techniques like search algorithms or artificial intelligence can help on maintaining the digital twin (e.g., changing certain parameters in order to consider the degradation of the real system). In addition, operational data can be complemented with further modelling techniques to propose changing subsystems that are out of the scope of testing interest by lighter (but still accurate) subsystems, reducing the time required by digital twins to be executed and permitting other novel testing methods. This leads us to the following open question:

\noindent\fbox{%
    \parbox{\textwidth}{%
        \textbf{Question 3:} Which additional test, verification and validation activities can be supported by multi-fidelity digital twins?
    }%
}

As previously mentioned, digital twins have supported several testing activities, at the different X-in-the-Loop test level, including test generation \cite{borg2020digital,menghi2020approximation,abdessalem2018testing} or anomalies detection \cite{Xu2021}. Our hypothesis has as presumes that multi-fidelity digital twins can support on significantly enhancing several test, verification and validation activities (e.g., test generation, test oracle generation, debugging) that are not supported with common digital twins. One problem with the X-in-the-Loop tests is that the test data are typically far from reality. One such high-gain activity that could be done by combining multi-fidelity digital twins with already established technology (e.g., mixed criticality hardware platforms and 5G communication technology), could be the test execution to be closer to operation, permitting this way the execution of more realistic tests. Ideally, when the software engineer has developed an algorithm, and some initial tests have been executed locally on the PC, the version could be committed and with the adequate orchestration algorithms, deployed along with the multi-fidelity digital twins in many real CPSs in operation. This would permit not only the use of real data, but also the execution of tests in parallel within many CPSs, testing the new version under many different conditions. This leads us to the following open questions:

\noindent\fbox{%
    \parbox{\textwidth}{%
        \textbf{Question 4:} How can we develop interoperable digital twins?
    }%
}

\noindent\fbox{%
    \parbox{\textwidth}{%
        \textbf{Question 5:} How can a multi-fidelity digital twin have the minimum impact on the resource-constrained characteristics of a CPS?
    }%
}

Both open questions are linked. In other software engineering areas (e.g., web development), containerized approaches (e.g., dockers) have emerged, increasing the interoperability of a piece of software when being executed. It is important to highlight, however, that CPSs are composed of embedded systems with limited resources, making it impossible to use this technology in practice. The interoperable solutions shall provide support to have minimum impact on the CPU, the memory and energy consumption. Furthermore, digital twins highly rely on their simulation environment (e.g., Simulink) and solver-specific configurations. For instance, a recent study showed that the same digital twin can produce different results for the same test scenario when being executed in two different environments \cite{borg2020digital}. In a similar line, if the digital twin is to be used locally, in parallel with the real execution of the CPS, this shall not be a high priority task in order not to interfere safety-critical functionalities. Instead, the multi-fidelity digital twin should be integrated within the non-critical tasks. This integration might require research on adapting operating systems and hypervisors of CPSs to accommodate this novel paradigm.

\section{Conclusion}

This paper presents the novel concept of multi-fidelity digital twins to enhance CPSs testing. We present our initial hypothesis and open questions that have remained unexplored so far. This paper represents the foundation research challenges for a novel CPSs testing approach that can bring significant benefits to CPS developers.

\section*{Acknowledgment} 
 Aitor Arrieta is part of the Software and Systems Engineering research group of Mondragon Unibertsitatea (IT1326-19), supported by the Department of Education, Universities and Research of the Basque Country. 


\bibliography{sample}
\bibliographystyle{elsarticle-num-names}







\end{document}